\documentclass[aps,prd,preprint,floatfix,12pt]{revtex4}
\usepackage{epsfig}
\usepackage{latexsym}
\usepackage{bm}

\begin{document}

\preprint{\rightline{ANL-HEP-CP-05-8}}

\title{Finite $dt$ dependence of the Binder cumulants for 3-flavor QCD at 
finite temperature and isospin density\footnote{Talk presented by J.~B.~Kogut
at the Workshop on QCD in Extreme Environments, Argonne National Laboratory,
29th June to 3rd July, 2004}}

\author{J.~B.~Kogut}
\address{Dept. of Physics, University of Illinois, 1110 West Green Street,
Urbana, IL 61801-3080, USA}
\author{D.~K.~Sinclair}
\address{HEP Division, Argonne National Laboratory, 9700 South Cass Avenue,
Argonne, IL 60439, USA}

\begin{abstract}
We simulate 3-flavour lattice QCD at small isospin chemical potential $\mu_I$
and finite temperature $T$. At $\mu_I=0$ there is a critical mass $m_c$ where
the finite-temperature transition changes from first order to a crossover. We
measure the $\mu_I$ dependence of the transition $\beta$ ($\beta_c$) for $m$
close to $m_c$. $\beta_c$ and hence $T_c$ decrease slowly with increasing
$\mu_I$. $\beta_c$ at finite $\mu_I$ is in good agreement $\beta_c$ at finite 
$\mu$ (quark-number chemical potential). We use fourth-order Binder cumulants
to determine the nature of this transition and to search for a critical
endpoint. We measure the $dt$ dependence of these cumulants and extrapolate to
$dt=0$. ($dt$ is the `time' increment used in the hybrid molecular-dynamics
simulations.) Preliminary measurements of these Binder cumulants show little
$\mu_I$ dependence. (Simulations at imaginary $\mu$ indicate that the $\mu$
dependence of the Binder cumulants is also weak.) This contrasts to the
$\mu_I$ dependence we observed at fixed $dt$.
\end{abstract}

\maketitle

\section{Introduction}

Although this is nominally the writeup of the talk delivered by JBK at the
Workshop on QCD in Extreme Environments, we have taken the liberty of updating
our results to reflect the `data' from simulations performed since then. We
have done this, since studies of the $dt$ dependence of observables indicate
that the Binder cumulants for the chiral condensate have a strong enough
dependence on $dt$ to change our conclusions. For our earlier results one
should see our contribution to Lattice 2004, Fermilab \cite{Kogut:2004qq}.

QCD at finite baryon/quark-number density and temperature is relevant to the
physics of relativistic heavy-ion collisions. The various accelerators which
probe this region of the phase diagram of nuclear matter include RHIC, CERN 
heavy-ion, AGS and SIS. Between them they should map out a considerable region
of this phase diagram.

QCD and in particular Lattice QCD, at a finite chemical potential $\mu$ for
quark-number, have a complex fermion determinant with a real part of indefinite
sign. For this reason standard simulation methods based on importance sampling
cannot be applied directly. Various methods have been developed to circumvent
this problem for small $\mu$ in the neighborhood of the finite temperature
transition \cite{Fodor:2001au,Fodor:2001pe,Fodor:2004nz,Allton:2002zi,
Karsch:2003va,deForcrand:2002ci,deForcrand:2003hx,D'Elia:2002gd,Azcoiti:2004rj}.
The transition $\beta$s and hence temperatures for finite $\mu$ and for finite
$\mu_I$ appear to be the same provided $\mu_I=2\mu$.  This equality was noted 
by the Bielefeld-Swansea group \cite{Allton:2002zi} and observed by us for 
the 2-flavor case \cite{Kogut:2004zg}. Thus the phase of the
determinant appears to be unimportant for determining the position of the
transition. We have argued that this was because the fluctuations in the
phase of the fermion determinant are moderate for small $\mu$ and high
temperatures, on the lattice sizes we use. We speculate that the nature of the
transition might also be insensitive to the phase of the determinant. If so,
simulations at finite $\mu_I$ offer an alternative way to study the finite
temperature transition at finite $\mu$. Lattice QCD at finite $\mu_I$ has a 
positive fermion determinant, so standard simulation methods work.

We are now extending our simulations to the 3-flavor case. Here we ignore the
complication that at finite $\mu_I$, this theory really describes QCD with
$3/2$ $u$ quark flavors and $3/2$ $d$ quark flavors. This theory can
be interpreted as 3-flavor QCD at finite $\mu=\mu_I/2$ where we use only the
magnitude of the fermion determinant and ignore the phase. 3 flavors are of
interest because they are closer to the physical $2+1$ flavors. 
In addition, for quark mass $m < m_c$, the phase transition is first order at
$\mu=\mu_I=0$ becoming a crossover for $m > m_c$
\cite{Aoki:1998gi,Karsch:2001nf,Christ:2003jk}. It is argued that the critical
point at $m=m_c$ moves continuously to higher quark mass as $\mu$ is
increased, where it becomes the sought-after critical endpoint
\cite{Karsch:2003va,deForcrand:2003hx}. Choosing a quark mass just above $m_c$,
one should be able to keep the critical endpoint as close to $\mu=0$ as
desired. In particular, one should be able to keep the critical $\mu$ in the
range where the $\mu$ and $\mu_I$ transitions are related. Thus we might
expect a critical endpoint at small $\mu_I$, which we could determine by
direct simulations.

Our initial simulations appeared to show evidence for this critical endpoint
\cite{Kogut:2004qq}. However, the Binder cumulants, which we use to determine
the nature of the transition, depend strongly on $dt$. We are currently
performing simulations to study this dependence to allow extrapolation to
$dt=0$. Indications are that the $\mu_I$ dependence of the $dt=0$ cumulants is
rather weak. This observation is consistent with what was predicted from
analytic continuation from imaginary $\mu$ by de Forcrand and Philipsen
\cite{deForcrand:2003hx}. However, our preliminary measurments favor a slow
increase with increasing $\mu_I$ while their measurements favor a slow
decrease with increasing $\mu$. Considerable more work will be needed to
identify the critical endpoint {\it if} it exists.

In section~2 we introduce our formulation of lattice QCD at finite $\mu_I$
and discuss its relationship to QCD at finite $\mu$. Section~3 describes our
current simulations of 3-flavor lattice QCD at finite $\mu_I$ and gives
preliminary results. In section~4 we give our conclusions and discuss what
remains to be done.

\section{QCD at finite $\mu_I$}

The staggered quark action for lattice QCD at finite isospin chemical potential
$\mu_I$ is \cite{Kogut:2004zg,Kogut:2002zg}
\begin{equation}
S_f=\sum_{sites} \left\{\bar{\chi}[D\!\!\!\!/(\frac{1}{2}\tau_3\mu_I)+m]\chi
                   + i\lambda\epsilon\bar{\chi}\tau_2\chi\right\},
\end{equation}
where $D\!\!\!\!/(\frac{1}{2}\tau_3\mu_I)$ is the normal staggered $D\!\!\!\!/$
for a staggered fermion field $\chi$ with 2 explicit flavor components, and 
$\bm{\tau}$ are the isospin Pauli spin matrices acting on this flavor space,
with the links in the $+t$ direction multiplied by 
$\exp(\frac{1}{2}\tau_3\mu_I)$ and those in the $-t$ direction multiplied by
$\exp(-\frac{1}{2}\tau_3\mu_I)$. Explicit $I_3$ symmetry breaking is provided
by the term proportional to $\lambda$. $\lambda \ne 0$ is only needed for
$\mu_I \geq m_\pi$ where, in the low temperature phase, $I_3$ is broken
spontaneously by a charged pion condensate.

The fermion determinant is 
\begin{equation}
\det \left\{ [D\!\!\!\!/(\frac{1}{2}\mu_I)+m]^\dagger
             [D\!\!\!\!/(\frac{1}{2}\mu_I)+m] + \lambda^2 \right\}.
\label{eqn:det}
\end{equation}
where in this case $D\!\!\!\!/$ is only a $1 \times 1$ matrix in isospin space,
which means that we only need use a single flavour-component fermion field in
our simulations. This determinant is real and positive allowing us to use
standard hybrid molecular-dynamics simulations with noisy fermions, to allow us
to tune the number of flavors from 8 down to $N_f$. Here we will choose $N_f=3$,
even though this represents $3/2$ isodoublets and probably does not have a
sensible continuum limit.

When $\lambda=0$, this determinant is the magnitude of the fermion determinant
at finite $\mu=\frac{1}{2}\mu_I$, so simulating at finite $\mu_I$ is
equivalent to simulating at finite $\mu$ using only the magnitude of the
determinant and neglecting the phase. Hence, the theory we are considering
corresponds to the 3-flavor theory at quark number chemical potential $\mu$,
in which the phase of the fermion determinant has been set to zero. We will
argue that for small enough $\mu$, the phase $\theta$ of the determinant is
well enough behaved, that it does not affect the position of the finite
temperature transition.

For simplicity let us consider a gluonic observable $X$. Its expectation value
at finite $\mu$ is given by the ratio
\begin{equation}
\langle X \rangle_\mu={\langle e^{i\theta} X \rangle_{\mu_I=2\mu}
                   \over      \langle e^{i\theta} \rangle_{\mu_I=2\mu}}
\end{equation}
where $\theta$ is the phase of the fermion determinant. If the denominator
\begin{equation}
\langle \cos \theta \rangle |_{\mu_I=2\mu}
\end{equation}
is reasonably large and varies smoothly as $\beta$ crosses the transition at
fixed $\mu_I$, i.e. it varies considerably more slowly than 
$\langle X \rangle$ for the volume considered, it will not have much effect
on the transition. We can therefore consider $e^{i\theta} X $ as our observable
and, to the extent to which the position of the transition is independent of
the observable, the finite $\mu$ transition and the finite $\mu_I$ transition
will be coincident. It is not unreasonable to assume that the nature of the
two transitions will also be the same. Because this argument only works for
finite volumes, it should be considered as a plausibility argument rather
than a proof.

\section{Simulations at small $\mu_I$ and finite temperature}

We are simulating the 3-flavor theory at finite temperature and small $\mu_I$,
i.e. $\mu_I < m_\pi$, since the transition to the phase where isospin ($I_3$)
is broken by a charged pion condensate occurs at $\mu_I = m_\pi$ for zero
temperature. Since no such condensed phase exists for QCD at finite $\mu$,
the physics at finite $\mu$ and that at finite $\mu_I$ can at best be related
for $\mu_I < m_\pi$ ($\mu < m_\pi/2$).

Our simulations are being performed on $N_t=4$ lattices where the 
zero-chemical-potential finite-temperature phase transition is first order
for $m < m_c(0)$ and a crossover for $m > m_c(0)$, with $m_c(0) \approx 0.033$.
At $m = m_c(0)$, the transition is a critical point in the universality class
of the 3-dimensional Ising model. As $\mu$ is increased, $m_c$ is expected to
increase so that the $m_c(\mu)$ becomes the critical endpoint. Fixing $m$ to
lie just above $m_c(0)$, the transition should be a crossover for $\mu < \mu_c$
with $\mu_c(m)$ the solution of $m=m_c(\mu_c)$. For $\mu > \mu_c$, the 
transition should become first order. The critical endpoint should again lie in
the Ising universality class. From the observation that the $\mu$ and $\mu_I$
transitions seem to be related, one might expect a corresponding endpoint for
finite $\mu_I$.

We are performing simulations at $m=0.03$, $m=0.035$ and $m=0.04$. Most of our
effort has been concentrated on $m=0.035$, just above $m_c(0)$, where we might
hope to find a critical endpoint for small $\mu_I$. Our simulations are being
performed on $8^3 \times 4$, $12^3 \times 4$ and $16^3 \times 4$ lattices.

\begin{figure}[htb]
\epsfxsize=6in
\centerline{\epsffile{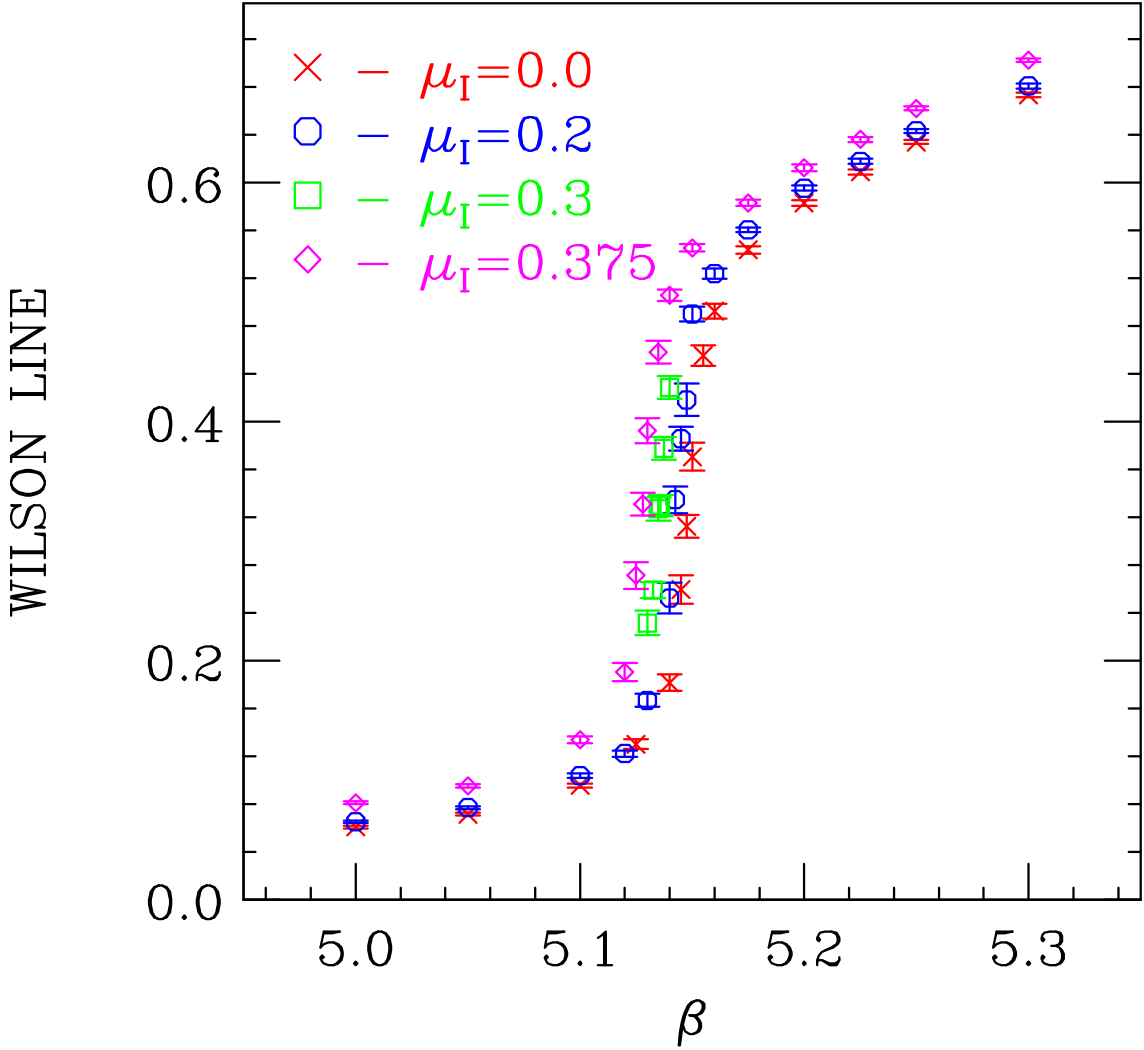}}
\caption{Wilson Lines as functions of $\beta$ for several values of $\mu_I$
on an $8^3 \times 4$ lattice with $m=0.035$.}\label{fig:wilson0.035}
\end{figure}

Figure~\ref{fig:wilson0.035} shows the Wilson line from our $8^3 \times 4$
simulations at $m=0.035$ for several values of $\mu_I < m_\pi$. The values of
$dt$ used for each of the $\mu_I$ values used in these simulations are
$dt=0.05$ for $\mu_I=0.0,0.2$, $dt=0.04$ at $\mu=0.25$, $dt=0.03125$ for 
$\mu_I=0.03$ and $dt=0.02$ at $\mu_I=0.0375$. The length of the simulations
for each of those $\beta$ values closest to the transition for each $\mu_I$ is
160,000 time-units(trajectories).

We study the nature of these transitions using 4-th order Binder cumulants 
\cite{binder}.  If $X$ is an observable, its Binder cumulant is defined by
\begin{equation}
B_4(X)={\langle (X - \langle X \rangle)^4 \rangle \over 
\langle (X - \langle X \rangle)^2 \rangle^2}
\end{equation}
The Binder cumulant of interest is that for the chiral condensate. Since we
only measure stochastic estimators of the chiral condensate, we need at least
4 independent stochastic estimates of $\langle\bar{\psi}\psi\rangle$ for each
configuration to obtain an unbiased estimate of $B_4$ -- we actually use 5
noise vectors for each configuration.

On a lattice of infinite spatial extent, $B_4(\langle\bar{\psi}\psi\rangle)$
would be $3$ for a crossover and $1$ for a first order transition, while for
a critical point in the Ising universality class it would have the value
$1.604(1)$. For finite lattices $B_4(\langle\bar{\psi}\psi\rangle)$ will lie
above the Ising value in the crossover region and below this value in the
first-order regime, passing smoothly through a value close to the Ising value
at the critical endpoint. As the size of the spatial lattice is increased, the
curve will pass more steeply through the Ising value. The critical endpoint
can be obtained by observing where the curves for different lattice sizes
cross, or where the curve for a given large lattice passes through the Ising
value. Ideally one should use several spatial volumes, observe where the
curves cross, and check that the value of $B_4$ at this point is (close to)
the Ising value.

Preliminary indications based on simulations with the $dt$ values given above
showed some evidence for such a critical endpoint. However, it was observed
that, while the value of $B_4$ fell with increasing $\mu_I$ at $\mu_I << m_\pi$,
it began to increase again long before $m_\pi$. This suggested that we should
study the $dt$ dependence of $B_4$. We are now performing a detailed finite 
$dt$ analysis. What we find is that the $dt$ dependence of the fluctuation
quantities -- $B_4(\langle\bar{\psi}\psi\rangle)$ and the susceptibility
$\chi(\langle\bar{\psi}\psi\rangle)$ -- is considerable, something we had
also observed in the 2 flavor case at $\mu_I > m_\pi$. The preliminary results
of this analysis which we present below indicate that considerable work will
be needed to determine whether there is evidence for a critical endpoint, and
if so, determine its position.

\begin{figure}[htb]
\epsfxsize=6in
\centerline{\epsffile{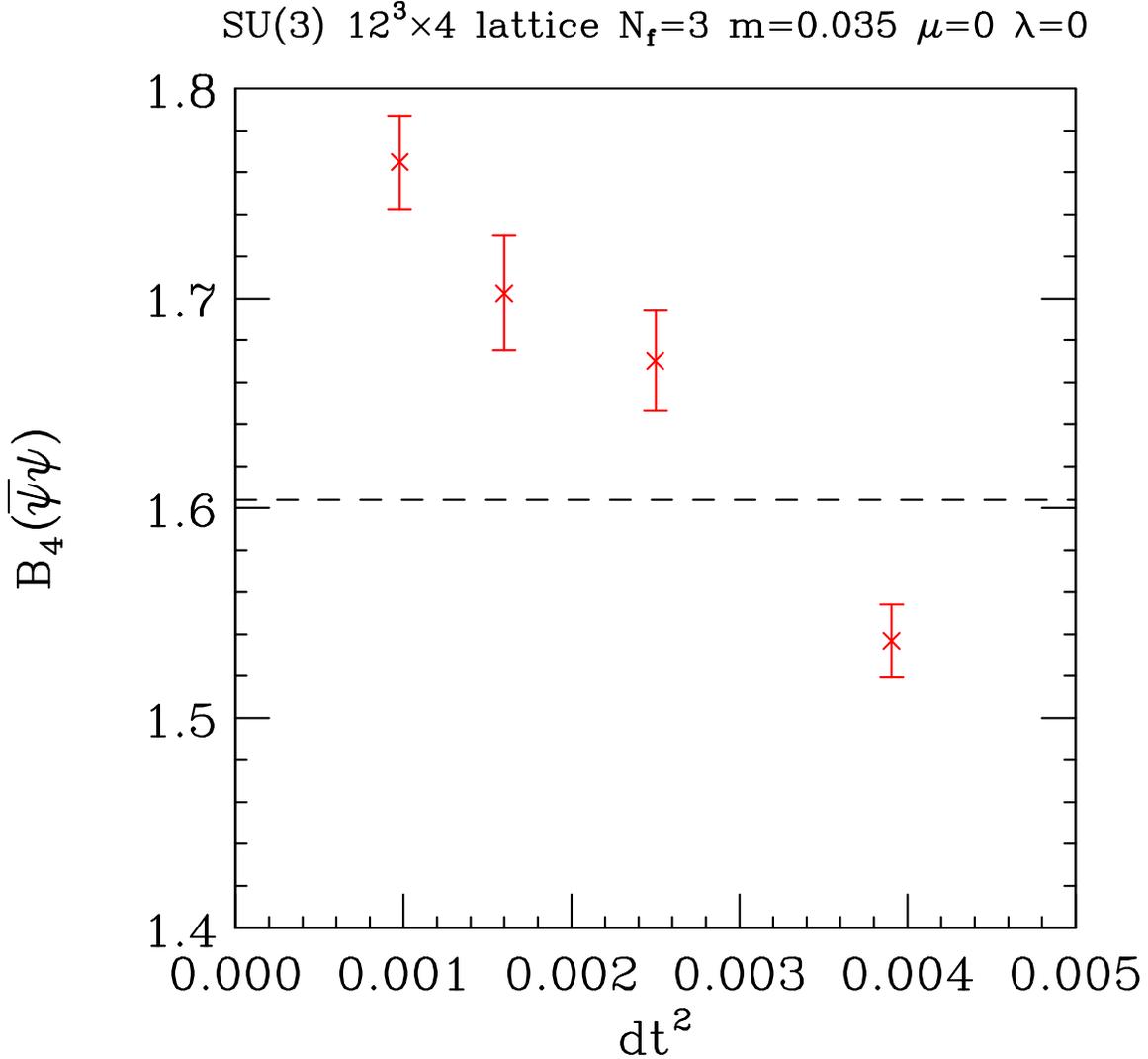}}
\caption{Estimates of $B_4(\langle\bar{\psi}\psi\rangle)$ as a function of
$dt^2$ at $m=0.035$ and $\mu_I=0$ on a $12^3 \times 4$ lattice.}
\label{fig:B4_mu0}
\end{figure}

Figure~\ref{fig:B4_mu0} shows the $dt$ dependence of $B_4$ for the chiral
condensate for $m=0.035$ and $\mu_I=0$ from our $12^3 \times 4$ runs at $dt$
values of $0.0625$, $0.05$, $0.04$ and $0.03125$. Each point is the value of
$B_4$ obtained from performing a Ferrenberg-Swendsen extrapolation 
\cite{Ferrenberg:yz} to its minimum from $\beta$ values close to the
transition. We have noted that the values of $\beta_c$ obtained from the
minima of $B_4$ and from the maxima of the chiral susceptibility are
consistent. We have averaged the results obtained from 2 or 3 different
$\beta$ values close to the transition, for each $dt$. The length of the runs
at each $\beta$ was 160,000 $\Delta t = 1$ trajectories. We also noted that
the values of $\beta_c$ obtained from different $\beta$s and the same $dt$
were consistent.
We note considerable variation of $B_4$ over the range of $dt$s used in these
runs. A fit linear in $dt^2$ gives $B_4(dt=0)=1.84(3)$ with a $\chi^2/dof$ of
$0.6$. This shows that, for $dt=0.0625$ the finite $dt$ error $\approx 16.5$\%.
Even at $dt=0.05$ which we used for our earlier simulations the error
$\approx 9$\%. This also shows that finite $dt$ errors can take one from 
$B_4 > 1.604$, which is normally interpreted as a crossover to $B_4 < 1.604$
which is commonly interpreted as indicative of a first order transition. 

\begin{figure}[htb]                                  
\epsfxsize=4in     
\centerline{\epsffile{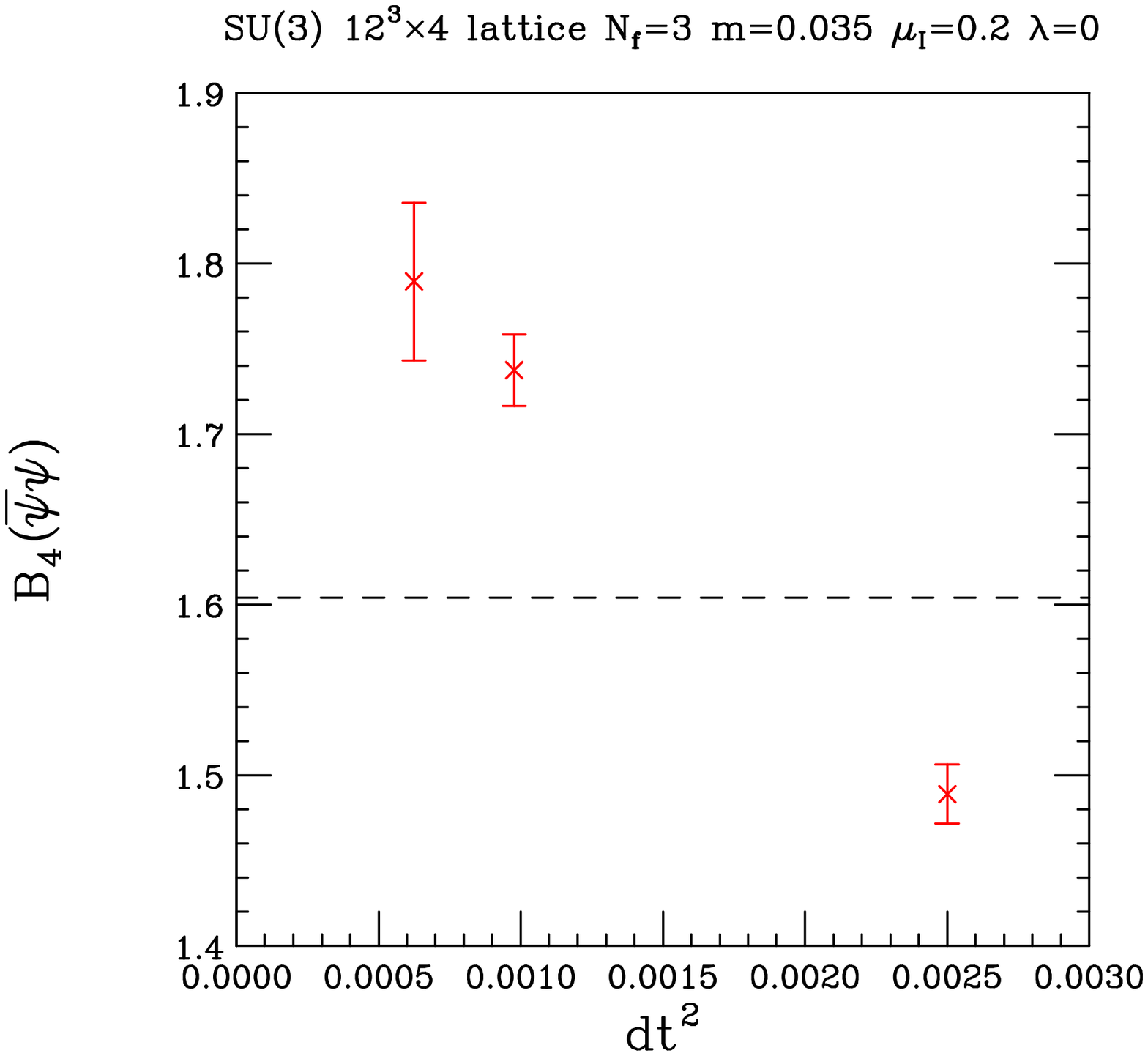}}
\vspace{0.2in}
\centerline{\epsffile{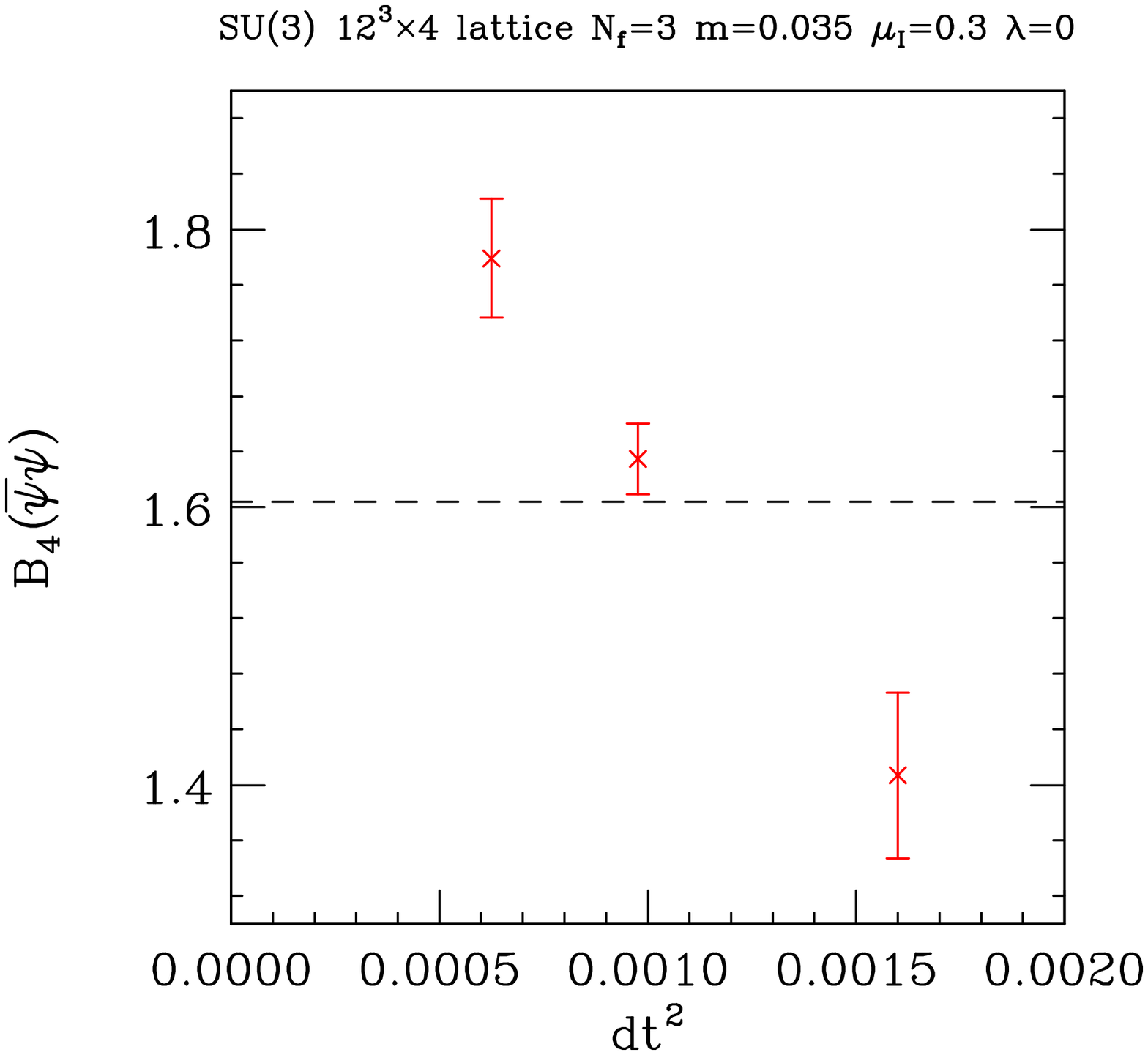}}
\caption{Estimates of $B_4(\langle\bar{\psi}\psi\rangle)$ as a function of
$dt^2$ at $m=0.035$ on a $12^3 \times 4$ lattice a) for $\mu_I=0.2$ and
b) for $\mu_I=0.3$.}
\label{fig:B4_mu>0}                                                        
\end{figure}

\begin{figure}[htb]
\epsfxsize=6in
\centerline{\epsffile{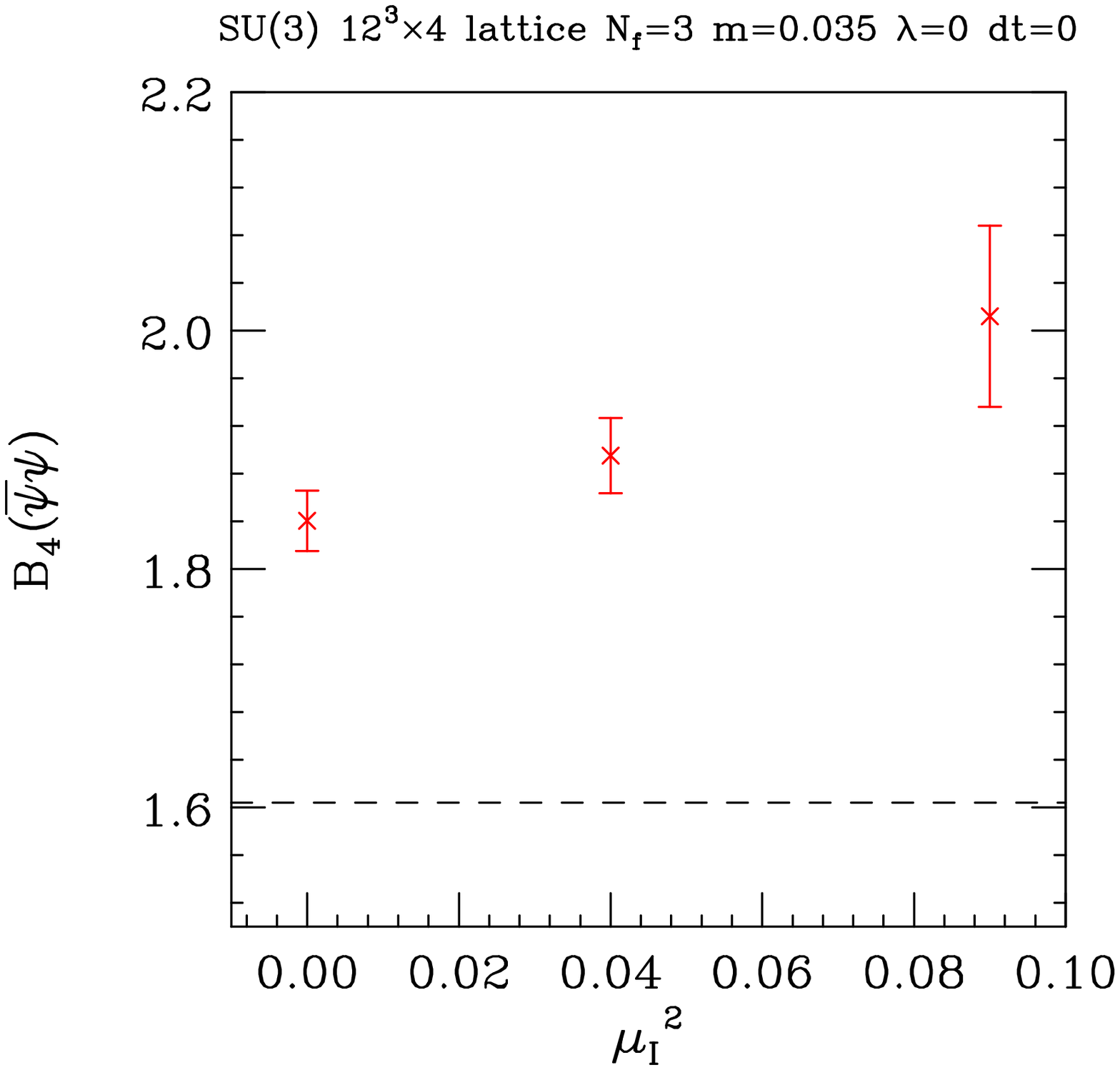}}
\caption{Estimates of $B_4(\langle\bar{\psi}\psi\rangle)$ extrapolated to 
$dt=0$, as functions of $\mu_I$ on a $12^3 \times 4$ lattice.}
\label{fig:B4dt=0}
\end{figure}

We are performing similar calculations of the finite $dt$ dependence of $B_4$
for the chiral condensate for $\mu_I > 0$. Preliminary results are shown in
figure~\ref{fig:B4_mu>0}.
Again we see a strong dependence of $B_4$ on $dt$. Because $\mu_I\ne 0$ makes
the Dirac operator more singular, smaller $dt$ values are required as $\mu_I$
is increased. An extrapolation linear in $dt^2$ to $dt=0$ gives $B_4=1.89(3)$
for $\mu_I=0.2$ and $B_4=2.01(8)$ for $\mu_I=0.3$. The rise in $B_4$ with
increasing $\mu_I$, although suggestive, is not statistically significant. All
we can claim at present is that there is little if any dependence of
$B_4(\langle\bar{\psi}\psi\rangle)$ on $\mu_I$ over this range. For
completeness we plot these extrapolated values of $B_4$ in
figure~\ref{fig:B4dt=0}.

\begin{figure}[htb]
\epsfxsize=6in
\centerline{\epsffile{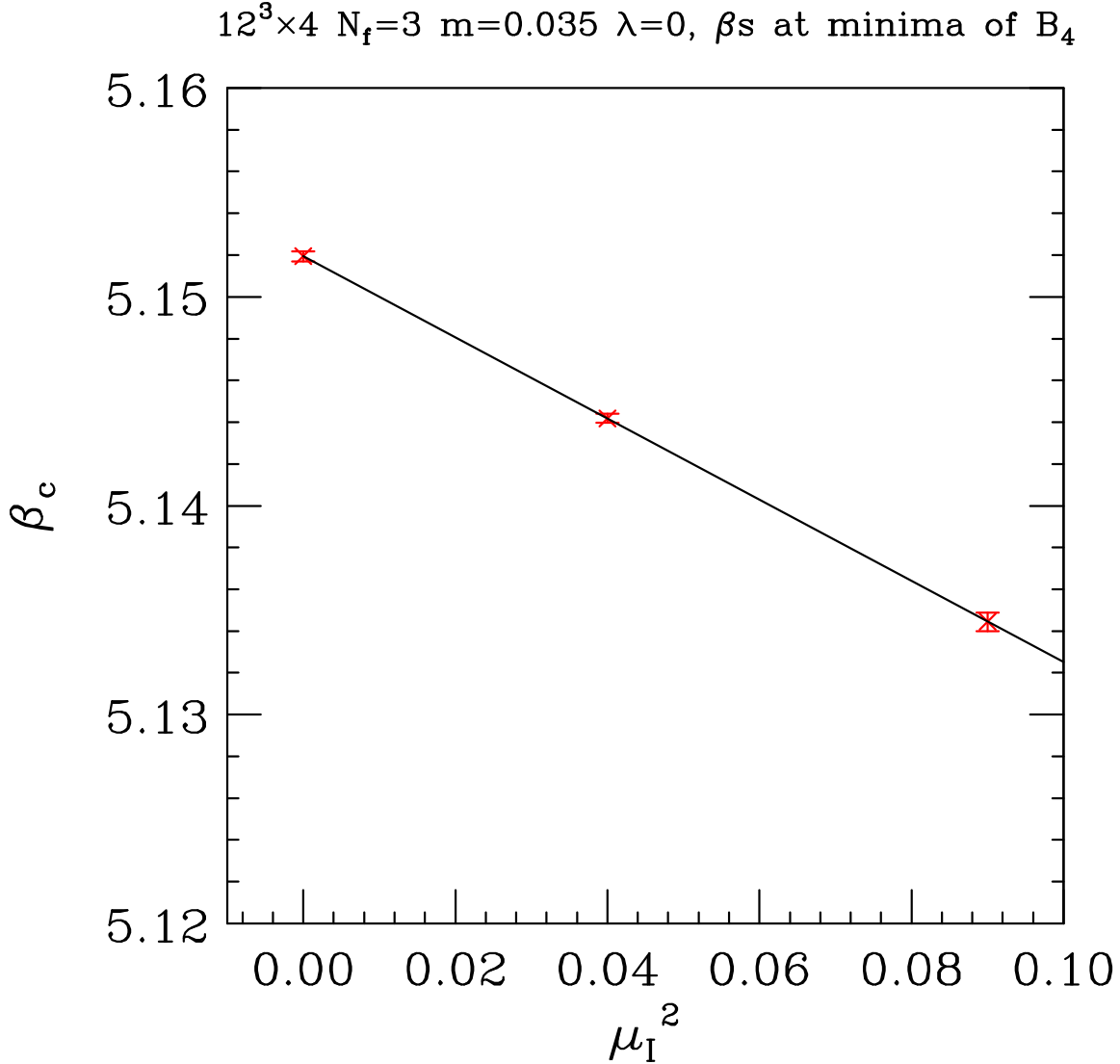}}
\caption{$\beta_c$ extrapolated to $dt=0$ as a function of $\mu_I$ on a
$12^3 \times 4$ lattice.}
\label{fig:beta_c}
\end{figure}

The values of $\beta_c$, the transition $\beta$ are calculated from the minima
of $B_4$. We have noted that these estimate of $\beta_c$ are in excellent
agreement with those obtained from the maxima of the susceptibilities for the
chiral condensate. Extrapolating these to $dt=0$, linearly in $dt^2$, we obtain
the results shown in figure~\ref{fig:beta_c}. The statistical errors are small,
but there are large systematics associated with this preliminary `data'. The
straight line fit shown in the figure is:
\begin{equation}
\beta_c = 5.15195 - 0.19428\:\mu_I^2.
\end{equation} 
We do not quote any errors because of the preliminary nature of this result.
If we interpret $\mu_I=2\mu$ as we have suggested, this is in good agreement
with the measurements of de Forcrand and Philipsen \cite{deForcrand:2003hx}.

\begin{figure}[htb]
\epsfxsize=6in
\centerline{\epsffile{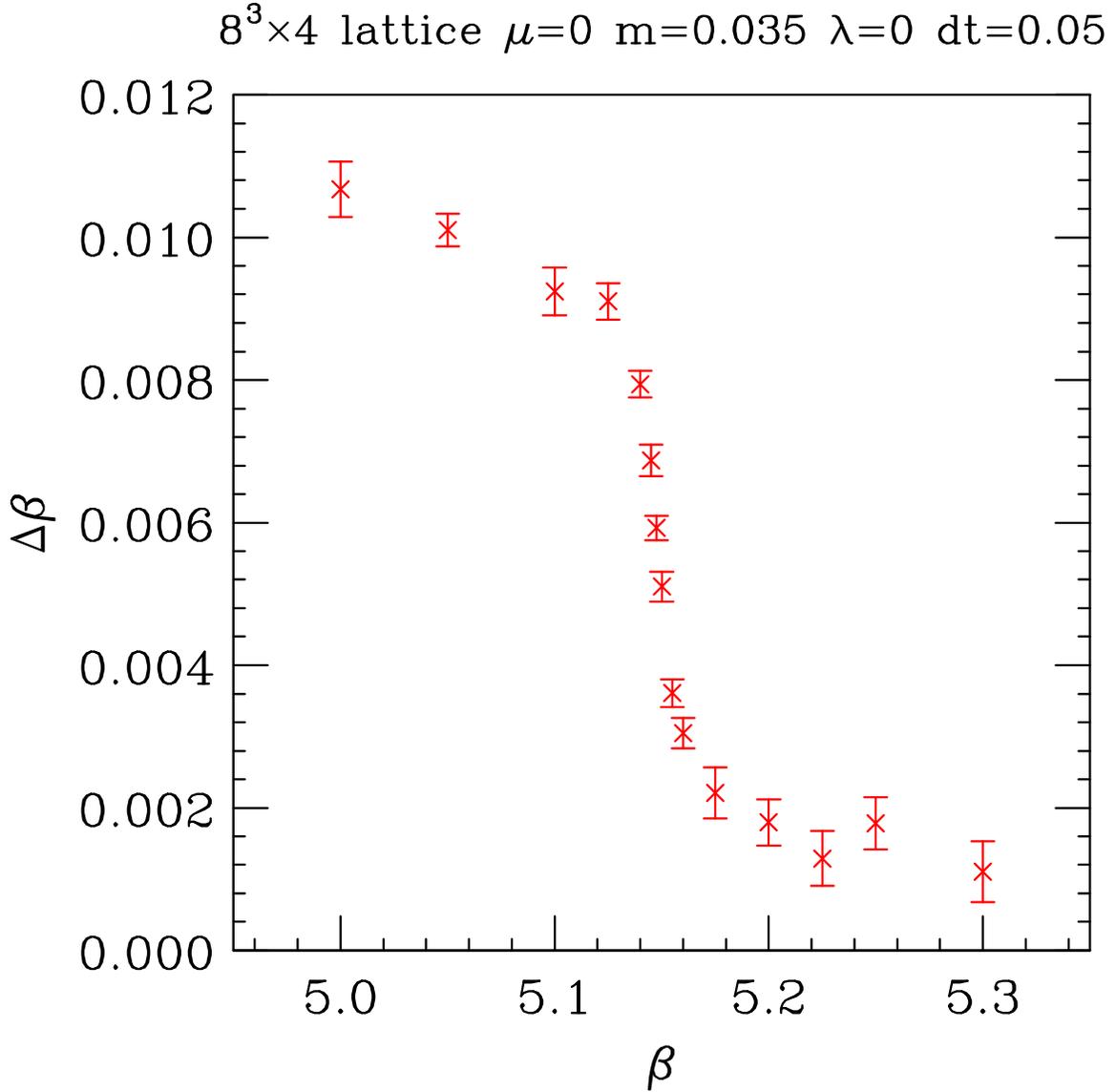}}
\caption{$\Delta\beta=\beta-\beta_{effective}$ as a function of $\beta$ at
$\mu_I=0$, $m=0.035$, $\lambda=0$ and $dt=0.05$ on an $8^3 \times 4$ lattice.}
\label{fig:BEFF}
\end{figure}

It is helpful to consider why the transition appears more like a first-order
transition as $dt$ is increased from zero. A brief discussion of this was
given in our 2-flavor paper \cite{Kogut:2004zg}. One of the effects of
simulating at a finite $dt$ using the hybrid molecular-dynamics method, which
is an inexact algorithm, is that the effective $\beta$ value
$\beta_{effective}$ measured from the kinetic term in the molecular-dynamics
Hamiltonian using the equipartition theorem is (on the average) less than the
input $\beta$, so that we are effectively running at a lower $\beta$ value.
The quantity $\Delta\beta=\beta-\beta_{effective}$ increases with increasing
$dt$. This would not be a problem if the shift were roughly constant as
$\beta$ was varied across the transition. However, this is not the case for a
constant $dt$. $\Delta\beta$ is significantly larger below the transition than
above. We illustrate this from our simulations at $\mu_I=0$, $m=0.035$,
$\lambda=0$ and $dt=0.05$ on an $8^3 \times 4$ lattice, where we have `data'
at a number of $\beta$ values over a range which includes the transition. This
is shown in figure~\ref{fig:BEFF}.

Now let us consider what the effect of such $\beta$ dependence for 
$\Delta\beta$ has on the behavior of the simulation close to the transition.
Here, a small change in $\beta$ causes a relatively large change in 
$\beta_{effective}$. This causes a relatively large change in all observables.
This, in turn, makes the transition appear more abrupt and hence nearer to
being first-order. Looking at this in another way, $\beta_{effective}$ needs
to undergo a relatively large fluctuation to take it from one side of the
transition to the other. This makes it more difficult for the system to
fluctuate between its low-$\beta$ domain and its high-$\beta$ domain. Such a
tendency to remain in one domain or the other, only `tunneling' between domains 
rather infrequently, is a property one associates with first-order behavior.
Hence the transition will appear to be closer to a first-order transition than
it really is. These considerations are in accord with our observations.

\section{Discussion and conclusions}

We are simulating 3-flavor lattice QCD with a chemical potential $\mu_I$ for
isospin ($I_3$), at small $\mu_I$ and temperatures close to the finite
temperature transition from hadronic/nuclear matter to a quark-gluon plasma.
These simulations are being performed on $8^3 \times 4$, $12^3 \times 4$, and
$16^3 \times 4$ lattices. In this regime, the behavior of QCD at finite $\mu_I$
and that at finite quark-number chemical potential $\mu$ appear similar for
$\mu_I=2\mu$. QCD at finite $\mu$ and QCD at finite $\mu_I$ are related if one
can neglect the phase angle of the determinant, or include it in the 
observables. The decrease in $\beta_c$ with increasing $\mu_I$ that we have
observed is consistent with that seen by de Forcrand and Philipsen if we make
the substitution $\mu=\frac{1}{2}\mu_I$, which adds further support to the
proposition that finite $\mu$ and finite $\mu_I$ are related.

At zero chemical potentials  there exists a critical mass $m_c$. For $m < m_c$
the finite temperature transition is first order. For $m > m_c$ this
transition is a mere crossover, with no real phase transition. At $m=m_c$ the
transition is a critical point in the universality class of the 3-dimensional
Ising model. For staggered lattices with $N_t=4$, $m_c \approx 0.033$. It has
been suggested that at finite $\mu$, the critical mass $m_c(\mu)$ moves
continuously (in $\mu$) to higher values, where it becomes the critical
endpoint. Hence the best place to look for a critical endpoint in $\mu$ or
$\mu_I$ should be for a quark mass just above $m_c$ where it would be expected
to be found at a small chemical potential. For this reason most of our
simulations have been performed at $m=0.035$.

We have used 4-th order Binder cumulants for the chiral condensate to determine
the nature of the finite temperature transition. A critical endpoint would be
where the Binder cumulants for different lattice sizes cross. At this point its
value should be close to the Ising value $1.604(1)$. Although our preliminary
results appeared to show evidence for such an endpoint, further simulations
studying the finite $dt$ dependence of this cumulant showed the finite $dt$
errors to be large at the $dt$s we were using. Our current estimates show
the extrapolated cumulants to lie well above the Ising value for the $\mu_I$
range being considered. There is no indication that these cumulants are falling
with increasing $\mu_I$. In fact our measurements show a slight rise with 
increasing $\mu_I$. However, this rise is not statistically significant,
especially considering that these preliminary measurements could have sizable
systematic errors. All that we can claim is that the $\mu_I$ dependence of
the Binder cumulants is very weak over this range. This is in qualitative
agreement with what was observed at finite $\mu$ by de Forcrand and Philipsen,
who observed a slight decrease in $B_4$ with increasing $\mu^2$, but could not
exclude a $B_4$ which is independent of $\mu^2$.

If we do not find evidence for a $B_4$ which decreases with increasing $\mu_I$
at this quark mass, we should check the situation even closer to $m_c$. The
lack of any critical end point, even for $m$ much closer to (but still above) 
$m_c$ would suggest that similar behavior would be found at small $\mu$, since
our estimates of the fluctuations of the phase angle as determined from
$\langle\cos\theta\rangle$ ($\theta$ is the phase of the fermion determinant
at $\mu=\frac{1}{2}\mu_I$) indicate that it is reasonably well behaved for
$m=0.035$ over most of the region $\mu_I < m_\pi$ for an $8^3 \times 4$ lattice.
This, in turn, would suggest that the critical endpoint at finite $\mu$ (if it
exists) is fluctuation induced. Thus we would expect to find it where 
fluctuations in $\theta$ become large, i.e. for $\mu \gtrsim m_\pi/2$. If so,
it would not move continuously away from $\mu=0$ as $m$ becomes larger than 
$m_c$, but might be expected to jump from $\mu=0$ to $\mu \approx m_\pi/2$,
as $m$ is increased beyond $m_c$.

The strong dependence of $B_4$ on $dt$ suggests that we should investigate the
use of recently formulated exact algorithms (see \cite{Kennedy:2004ae} and
references contained therein), for our simulations. Since such methods rely on
stochastic approximations to the fermion determinant, they could also help us
in applying reweighting by the phase of the fermion determinant as a way of
studying finite $\mu$ in a quantitative fashion, to obtain the
equation-of-state. Here we note recent work indicating that finite $\mu_I$
simulations provide a better platform for reweighting than reweighting from
simulations at $\mu=0$ \cite{deForcrand:2002pa}.

\section*{Acknowledgements}

JBK is supported in part by the National Science foundation under grant
grant NSF PHY03-04252. DKS is supported by the US Department of Energy under
contract W-31-109-ENG-38. The simulations reported here are being run on the 
Jazz Linux cluster belonging to the LCRC at Argonne National Laboratory, the
Tungsten Linux cluster at NCSA through an NRAC allocation and on Linux PCs in
the High Energy Physics Division at Argonne National Laboratory. Archival
storage is provided on the HPSS at NERSC. We thank P.~de Forcrand, S.~Hands,
S.~Ejiri, F.~Karsch and O.~Philipsen for helpful discussions.

\end{document}